\documentclass[12pt]{iopart}
\usepackage{iopams}
\eqnobysec

\begin{document}


\def\eqref#1{(\ref{#1})}
\def\eqrefs#1#2{(\ref{#1}) and~(\ref{#2})}
\def\eqsref#1#2{(\ref{#1}) to~(\ref{#2})}
\def\sysref#1#2{(\ref{#1}),~(\ref{#2})}

\def\Eqref#1{Eq.~(\ref{#1})}
\def\Eqrefs#1#2{Eqs.~(\ref{#1}) and~(\ref{#2})}
\def\Eqsref#1#2{Eqs.~(\ref{#1}) to~(\ref{#2})}
\def\Sysref#1#2{Eqs. (\ref{#1}),~(\ref{#2})}

\def\secref#1{Sec.~\ref{#1}}
\def\secrefs#1#2{Sec.~\ref{#1} and~\ref{#2}}

\def\appref#1{Appendix~\ref{#1}}

\def\Ref#1{Ref.~\cite{#1}}

\def\Cite#1{${\mathstrut}^{\cite{#1}}$}

\def\tableref#1{Table~\ref{#1}}

\def\figref#1{Fig.~\ref{#1}}

\hyphenation{Eq Eqs Sec App Ref Fig}

\def\EQ{\begin{equation}}
\def\EQs{\begin{eqnarray}}
\def\endEQ{\end{equation}}
\def\endEQs{\end{eqnarray}}


\def\fewquad{\qquad\qquad}
\def\severalquad{\qquad\fewquad}
\def\manyquad{\qquad\severalquad}
\def\manymanyquad{\manyquad\manyquad}

\def\downupindices#1#2{{}^{}_{#1}{}_{}^{#2}}
\def\updownindices#1#2{{}_{}^{#1}{}^{}_{#2}}
\def\mixedindices#1#2{{\mathstrut}^{#1}_{#2}}
\def\downindex#1{{}_{#1}}
\def\upindex#1{{}^{#1}}

\def\eqtext#1{\hbox{\rm{#1}}}

\def\hp#1{\hphantom{#1}}

\def\Parder#1#2{
\mathchoice{\partial{#1} \over\partial{#2}}{\partial{#1}/\partial{#2}}{}{} }
\def\nParder#1#2#3{
\mathchoice{\partial^{#1}{#2} \over\partial{#3}^{#1}}{\partial^{#1}{#2}/\partial{#3}^{#1}}{}{} }
\def\mixedParder#1#2#3#4{
\mathchoice{\partial^{#1}{#2} \over\partial{#3}\partial{#4}}{\partial^{#1}{#2}/\partial{#3}\partial{#4}}{}{} }
\def\parder#1#2{\partial{#1}/\partial{#2}}
\def\nparder#1#2{\partial^{#1}/(\partial {#2})^{#1}}
\def\mixedparder#1#2#3#4{\partial^{#1}{#2}/\partial{#3}\partial{#4}}
\def\parderop#1{\partial/\partial{#1}}
\def\mixedparderop#1#2#3{\partial^{#1}/\partial{#2}\partial{#3}}

\def\x#1#2{x^{#1}_{#2}}
\def\u#1{u\upindex{#1}}
\def\ujet#1{\nder{}{#1}u}
\def\uder#1#2{u\upindex{#1}\downindex{,#2}}
\def\deru#1#2{\der{\u{#1}{#2}}}
\def\Du#1{u\downindex{#1}}

\def\parderu#1#2#3#4{\mixedParder{#2}{\u{#1}(x)}{#3\cdots}{#4}}

\def\ueq#1#2{\Upsilon\mixedindices{#1}{#2}}
\def\lin#1#2#3#4{{\cal L}_{#1}({#2})\mixedindices{#3}{#4}}
\def\adlin#1#2#3{{\cal L}^*_{#1}({#2})\downindex{#3}}
\def\linop#1{{\cal L}_{#1}}
\def\adlinop#1{{\cal L}^*_{#1}}
\def\symm#1#2{\eta\mixedindices{#2}{#1}}
\def\adsymm#1{\omega\downindex{#1}}
\def\Q#1#2{Q\mixedindices{#1}{#2}}
\def\scsymm#1#2{\eta_{\rm s}\mixedindices{#2}{#1}}
\def\scvar{\delta_{\rm s}}

\def\Qseq#1{\Q{(#1)}}

\def\ueqder#1#2#3{\ueq{#3}{}\mixedindices{\hp{,}#2}{,#1}}

\def\P#1#2{\Phi\mixedindices{#1}{#2}} 
\def\curr#1#2{\Psi\mixedindices{#1}{#2}} 
\def\triv#1{\Theta\upindex{#1}} 

\def\der#1{\partial_{#1}}
\def\nder#1#2{\partial^{#2}_{#1}}
\def\D#1{D\downindex{#1}}
\def\nD#1#2{D\mixedindices{#2}{#1}}
\def\Dt#1{D\mixedindices{#1}{t}}
\def\Dx#1{D\mixedindices{#1}{x}}
\def\coder#1{\partial^{#1}}
\def\coD#1{D\upindex{#1}}

\def\p#1{{p^{(#1)}}}
\def\q#1{{q^{(#1)}}}
\def\r#1{{r_{(#1)}}}
\def\s#1{{s^{(#1)}}}

\def\w#1{w_{#1}}

\def\n#1#2{n\mixedindices{#1}{#2}}

\def\lieder#1{{\cal L}_{#1}}
\def\kv#1#2{\xi\mixedindices{#1}{#2}}
\def\veckv{\vec \xi}

\def\id#1#2{\delta\mixedindices{#2}{#1}}
\def\inv{{}^{-1}}


\def\invDx{D\mixedindices{-1}{x}}
\def\R{{\cal R}}
\def\adR{\R^*}
\def\nlR{\R_{\rm nonloc}}
\def\nladR{\adR_{\rm nonloc}}
\def\PR{{\hat{\cal R}}}

\def\intprod{\,\lrcorner }


\def\vecu{{\vec u}}
\def\vecx{{\vec x}}
\def\vecder{\der{\vecx}}
\def\vecueq{{\vec \Upsilon}}
\def\vecsymm#1{{\vec \eta}\upindex{#1}}
\def\vecadsymm#1{{\vec \omega}\downindex{#1}}
\def\vecscsymm{{\vec \eta}\downindex{\rm s}}
\def\Dvecu#1{\vecu\downindex{#1}}

\def\v#1{v\upindex{#1}}
\def\vecv{{\vec v}}

\def\invdens{\rho^{-1}}
\def\dens{\rho}

\def\vort#1{\Omega\upindex{#1}}
\def\vecvort{{\vec \Omega}}

\def\cross#1#2{\epsilon\updownindices{#1}{#2}} 
\def\grad{{\vec \partial}}


\def\T#1#2{T\updownindices{#1}{#2}}

\def\pow{\upsilon}
\def\div{{\rm div}}
\def\bdkv{{\kv{}{}}}


\def\A#1#2{A\mixedindices{#1}{#2}} 
\def\F#1#2{F\mixedindices{#1}{#2}} 
\def\Ajet#1{\nder{}{#1}A}

\def\sf#1{\chi\upindex{#1}}

\def\ymder#1{\nabla\mixedindices{A}{#1}}
\def\ymcoder#1{\nabla\upindex{A#1}}
\def\c#1#2{c\upindex{#1}\downindex{#2}} 
\def\j#1#2{j\updownindices{#1}{#2}} 
\def\ck#1#2{k\mixedindices{#1}{#2}} 

\def\liealg{{\cal G}}

\def\jT#1#2{{\tilde T}\updownindices{#1}{#2}}


\def\g#1#2{g\mixedindices{#1}{#2}} 
\def\gjet#1{\nder{}{#1}g}
\def\G#1#2{G\downupindices{#1}{#2}}
\def\curv#1#2{R\downupindices{#1}{#2}}
\def\vol#1#2{\epsilon\mixedindices{#2}{#1}} 

\def\covder#1{\nabla\mixedindices{g}{#1}}
\def\covcoder#1{\nabla\upindex{g #1}}

\def\trrvsymm#1{\bar\eta\upindex{#1}}
\def\coQ#1{Q\upindex{#1}}

\def\vf#1#2{\zeta\mixedindices{#1}{#2}}
\def\scvf#1#2{\xi\mixedindices{#1}{#2}}

\def\twist{\upsilon}
\def\dualkv#1#2{\chi\mixedindices{#1}{#2}}


\def\ie/{i.e.}
\def\eg/{e.g.}
\def\etc/{etc.}
\def\const{{\rm const}}

\def\Rnum{{\mathbb R}}

\title{Conservation laws of scaling-invariant field equations }

\author{ Stephen C. Anco }

\address{
Department of Mathematics\\
Brock University, 
St. Catharines, ON Canada }

\ead{sanco@brocku.ca}

\begin{abstract}
A simple conservation law formula for field equations 
with a scaling symmetry is presented. 
The formula uses adjoint-symmetries of the given field equation
and directly generates all local conservation laws
for any conserved quantities having non-zero scaling weight. 
Applications to several soliton equations, 
fluid flow and nonlinear wave equations, 
Yang-Mills equations and the Einstein gravitational field equations
are considered. 
\end{abstract}


\maketitle

\section{ Introduction }

Conservation laws are central to the analysis of physical field equations
by providing conserved quantities, 
such as energy, momentum, and angular momentum. 
For a given field equation, 
local conservation laws are well-known to arise through multipliers
\cite{Olver-book},
analogous to integrating factors of ODEs \cite{ourbook}, 
with the product of the multiplier and the field equation 
being a total divergence expression. 
Such divergences correspond to a conserved current vector
for solutions of the field equation
whenever the multiplier is non-singular. 
If a field equation possesses a Lagrangian, 
Noether's theorem \cite{Olver-book}
shows that the multipliers for local conservation laws
consist of symmetries of the field equation such that 
the action principle is invariant (to within a boundary term). 
Moreover, the variational relation between 
the Lagrangian and the field equation 
yields an explicit formula for the resulting conserved current vector. 
This characterization of multipliers for a Lagrangian field equation
has a generalization to any field equation by means of adjoint-symmetries
\cite{Anco-Bluman1,Anco-Bluman2}, 
whether or not a Lagrangian formulation exists. 

Recall, geometrically, symmetries are tangent vector fields 
on the solution space of a field equation
and thus are determined as field variations satisfying 
the linearization of the field equation on its entire solution space. 
Adjoint-symmetries are defined to satisfy the adjoint equation of 
the symmetry determining equation 
on the solution space of a field equation \cite{trivial}. 
(As such, unlike for symmetries, there is no obvious geometrical motion
or invariance associated with adjoint-symmetries.)
Through standard results in the calculus of variations \cite{Olver-book}, 
it is known that 
the multipliers for local conservation laws are precisely 
adjoint-symmetries of the field equation 
subject to a certain adjoint invariance condition
\cite{Anco-Bluman1,invcondition}. 
This allows a system of determining equations for multipliers 
to be formulated in terms of the adjoint-symmetry determining equation
augmented by extra determining equations \cite{Anco-Bluman2,deteq}. 
In addition, the resulting conservation laws 
are yielded by means of a homotopy integral expression 
\cite{Olver-book,Anco-Bluman1,Anco-Bluman2}
involving just the field equation and the multiplier,
which is derived from the adjoint invariance condition,
analogously to the line integral formula for first integrals of ODEs
\cite{ourbook}. 

The purpose of this paper is to show that 
in the physically interesting situation 
where a field equation possesses a scaling symmetry, 
then the adjoint invariance condition and homotopy integral formula
can be completely by-passed for obtaining conservation laws. 
In particular, a simple algebraic formula that directly generates 
conservation laws in terms of adjoint-symmetries for any such field equation 
is presented. 
Most important, when applied to a multiplier, 
the formula recovers the corresponding conservation law 
determined by the multiplier, 
to within a proportionality factor. 
This factor turns out to be the scaling weight of the conserved quantity
defined from the conservation law. 
Consequently, all conserved quantities with non-zero scaling weight
are obtainable from this formula \cite{linear}. 

In \secref{formula}, the conservation law formula is derived.
Examples and applications of this formula are presented in \secref{examples}.
As new results, 
first, 
a recursion formula is obtained for the local higher-order conservation laws
of the sine-Gordon equation 
and a vector generalization of the Korteweg-de Vries equation; 
second, a simple proof is given for closing a gap in the classification of
local conservation laws of the Yang-Mills equations
and Einstein gravity equations. 
Some concluding remarks are made in \secref{conclude}.

\section{ Conservation Law Formula }
\label{formula}

Consider a general system of field equations 
\EQ\label{ueq}
\ueq{A}{}(x,u,\ujet{},\ujet{2},\ldots) =0
\endEQ
for field variables $\u{a}(x)$
depending on a total of $n\geq 2$ time and space variables $\x{\alpha}{}$, 
with $\ujet{k}$ denoting partial derivatives
$\uder{a}{\alpha_1\cdots\alpha_k}
=\parderu{a}{k}{\x{\alpha_1}{}}{\x{\alpha_k}{}}$,
up to some finite differential order. 
(The coordinate indices $\alpha,\beta,\gamma$ run $0$ to $n-1$;
the field index $a$ runs $1$ to $N$;
the equation index $A$ runs $1$ to $m$.
Summation is assumed over any repeated indices. 
This formalism allows the number of components of 
the fields $\u{a}$ and equations $\ueq{A}{}$ to be different.)
For simplicity of presentation,
the differential order of system \eqref{ueq} 
will be restricted to $k\leq 2$. 

Symmetries of the field equations \eqref{ueq} are the solutions 
$\delta\u{a}=\symm{}{a}(x,u,\ujet{},\ujet{2},\ldots)$ 
of the linearized equations 
\EQ\label{usymmeq}
\lin{\ueq{}{}}{\symm{}{}}{A}{}
= \symm{}{a}\ueqder{a}{}{A} +(\D{\alpha}\symm{}{a})\ueqder{a}{\alpha}{A} 
+(\D{\alpha}\D{\beta}\symm{}{a})\ueqder{a}{\alpha\beta}{A} 
=0
\endEQ
for all $\u{a}(x)$ satisfying system \eqref{ueq},
where $\D{\alpha}$ is the total derivative with respect to $\x{\alpha}{}$,
and where $\ueqder{a}{}{A}$, $\ueqder{a}{\alpha}{A}$, \etc/
denote partial derivatives 
$\partial\ueq{A}{}/\partial\u{a}$, 
$\partial\ueq{A}{}/\partial\uder{a}{\alpha}$, \etc/.
The adjoint of equation \eqref{usymmeq} is given by 
\EQ\label{uadsymmeq}
\adlin{\ueq{}{}}{\adsymm{}}{a} 
= \adsymm{A}\ueqder{a}{}{A} -\D{\alpha}( \adsymm{A} \ueqder{a}{\alpha}{A} )
+\D{\alpha}\D{\beta}( \adsymm{A} \ueqder{a}{\alpha\beta}{A} )
=0
\endEQ
whose solutions $\adsymm{A}(x,u,\ujet{},\ujet{2},\ldots)$ 
for all $\u{a}(x)$ satisfying system \eqref{ueq}
are the adjoint-symmetries of the field equations \eqref{ueq}.
Note the operators $\linop{\ueq{}{}}$ and $\adlinop{\ueq{}{}}$
are related by the identity 
\EQ\label{mainid}
\adsymm{A} \lin{\ueq{}{}}{\symm{}{}}{A}{}
- \symm{}{a} \adlin{\ueq{}{}}{\adsymm{}}{a} 
= \D{\alpha} \P{\alpha}{}(\adsymm{},\symm{}{};\ueq{}{})
\endEQ
with
\EQ\label{curr}
\P{\alpha}{}(\adsymm{},\symm{}{};\ueq{}{})
= \symm{}{a}( \adsymm{A}\ueqder{a}{\alpha}{A} 
-\D{\beta}( \adsymm{A}\ueqder{a}{\alpha\beta}{A} ) )
- (\D{\beta}\symm{}{a}) \adsymm{A}\ueqder{a}{\alpha\beta}{A} . 
\endEQ
Hence, this expression \eqref{curr} yields a local conservation law
$\D{\alpha} \P{\alpha}{}(\adsymm{},\symm{}{}) =0$
for any pair $\adsymm{A},\symm{}{a}$, 
on all solutions of the field equations \eqref{ueq}.

Now suppose the field equations are invariant under a scaling of 
the variables
\EQ\label{scaling}
\x{\alpha}{} \rightarrow \lambda^\p{\alpha} \x{\alpha}{} ,\quad
\u{a} \rightarrow \lambda^\q{a} \u{a} , 
\endEQ
with $\p{\alpha}=\const$, $\q{a}=\const$.
From the corresponding scaling symmetry, given by 
\EQ\label{scsymm}
\scvar\u{a}=\scsymm{}{a}(x,u,\ujet{})
= \q{a} \u{a} -\p{\alpha} \x{\alpha}{}\uder{a}{\alpha} ,
\endEQ
the expression $\P{\alpha}{}(\adsymm{},\scsymm{}{})$ 
produces a conserved current in terms of 
any adjoint-symmetry $\adsymm{A}$.

{\bf Proposition 2.1}:
For scaling invariant field equations \eqref{ueq}, 
every adjoint-symmetry \eqref{uadsymmeq} generates
a conserved current on all solutions of \eqref{ueq}
by the formula
\EQs
\P{\alpha}{\adsymm{}} = &&
( \q{a} \u{a} -\p{\gamma}\x{\gamma}{}\uder{a}{\gamma} )
( \adsymm{A}\ueqder{a}{,\alpha}{A} 
-\D{\beta}( \adsymm{A}\ueqder{a}{,\alpha\beta}{A} ) )
\nonumber\\&& 
+( (\p{\beta}-\q{a}) \uder{a}{\beta} 
+\p{\gamma}\x{\gamma}{}\uder{a}{\beta\gamma} ) 
\adsymm{A}\ueqder{a}{\alpha\beta}{A} . 
\label{currformula}
\endEQs

Consider, now, a multiplier $\Q{}{A}(x,u,\ujet{},\ujet{2},\ldots)$
for a local conservation law 
\EQ\label{conslaw}
\Q{}{A}\ueq{A}{} =\D{\alpha} \curr{\alpha}{\Q{}{}}
\endEQ
that is assumed to be homogeneous \cite{homogeneous}
under the scaling symmetry, so
\EQ\label{Qscaling}
\scvar\Q{}{A}= \lin{\Q{}{}}{\scsymm{}{}}{}{A} 
= \r{A} \Q{}{A} -\p{\alpha} \x{\alpha}{}\D{\alpha} \Q{}{A} 
\endEQ
with scaling weight $\r{A}=\const$. 
Let $\s{A}=\const$ be the scaling weight of the field equations,
\EQ\label{eqscaling}
\scvar\ueq{A}{}= \lin{\ueq{}{}}{\scsymm{}{}}{A}{} 
= \s{A} \ueq{A}{} -\p{\alpha} \x{\alpha}{}\D{\alpha} \ueq{A}{} . 
\endEQ
Due to the scaling homogeneity of $\curr{\alpha}{\Q{}{}}$, 
the constant $\r{A}+\s{A}$ is independent of the index $A$.
Then, the following important relation holds between
the conserved currents $\curr{\alpha}{\Q{}{}}$ and $\P{\alpha}{\Q{}{}}$. 

{\bf Theorem 2.2}: 
In terms of the scaling weights of 
the field equations \eqref{eqscaling} and the multiplier \eqref{Qscaling}, 
every local conservation law \eqref{conslaw} satisfies 
the scaling relation 
\EQ\label{scalingformula}
\P{\alpha}{\Q{}{}} \simeq \w{\Q{}{}} \curr{\alpha}{\Q{}{}} ,\quad
\w{\Q{}{}} = \r{A}+\s{A} +\sum_\alpha \p{\alpha}
\endEQ
for all $\u{a}(x)$ satisfying the field equations,
where ``$\simeq$'' denotes equality to within a trivial conserved current
\cite{conslaw}
$\D{\beta} \triv{\alpha\beta}$
for some local expression $\triv{\alpha\beta}=-\triv{\beta\alpha}$. 
Moreover, 
$\w{\Q{}{}}$ is simply the scaling weight of the flux integral
$\int \curr{\alpha}{\Q{}{}} \n{}{\alpha} \rmd^{n-1}x$
defined on any $(n-1)$-dimensional hypersurface 
$\x{\alpha}{}\n{}{\alpha} =\const$
(with normal vector $\n{}{\alpha}$),
\EQ\label{scalingweight}
\scvar \int \curr{\alpha}{\Q{}{}} \n{}{\alpha} \rmd^{n-1}x 
= \w{\Q{}{}} \int \curr{\alpha}{\Q{}{}} \n{}{\alpha} \rmd^{n-1}x . 
\endEQ

{\bf Definition 2.3}: 
A conservation law \eqref{conslaw} will be called (non)critical
with respect to the scaling \eqref{scaling}
if the scaling weight \eqref{scalingweight} of the corresponding 
conserved quantity is (non)zero. 

{\bf Corollary 2.4}: 
As all multipliers necessarily are given by adjoint-symmetries,
the conservation law formula \eqref{currformula}
consequently generates all noncritical conservation laws of
the field equations \eqref{ueq}. 

The proof of relation \eqref{scalingformula} 
starts from the identity \eqref{mainid}
with $\symm{}{a}=\scsymm{}{a}$. 
We substitute the condition 
$\adlin{\adsymm{}}{\ueq{}{}}{a} = - \adlin{\ueq{}{}}{\adsymm{}}{a}$
on $\adsymm{A}$
(holding for all $\u{a}(x)$ without use of the field equations), 
which is necessary and sufficient \cite{Olver-book}
for an adjoint-symmetry to be a multiplier $\Q{}{A}=\adsymm{A}$. 
Here, $\adlinop{\adsymm{}}$ is the adjoint of the linearization operator
$\linop{\adsymm{}}$ defined analogously to 
$\adlinop{\ueq{}{}}$ and $\linop{\ueq{}{}}$.
We next use the adjoint relation 
\EQ
\scsymm{}{a}\adlin{\adsymm{}}{\ueq{}{}}{a} 
= \ueq{A}{}\lin{\adsymm{}}{\scsymm{}{}}{}{A} 
- \D{\alpha} \curr{\alpha}{}(\ueq{}{},\scsymm{}{};\adsymm{}) .
\endEQ
Substituting the scaling relations \eqrefs{Qscaling}{eqscaling}, 
followed by integrating by parts, we obtain
\EQ
\w{\Q{}{}} \Q{}{A}\ueq{A}{} 
= \D{\alpha}( \P{\alpha}{\Q{}{}} 
+ \curr{\alpha}{}(\ueq{}{},\scsymm{}{};\Q{}{}) )
\endEQ
where, note, the last term in this divergence vanishes when $\ueq{A}{}=0$. 
Then conservation law equation \eqref{conslaw} 
leads to the scaling relation \eqref{scalingformula}.

\section{ Examples and Applications }
\label{examples}

\subsection{ Soliton equations }

For applications of the main conservation law formulas
\eqrefs{currformula}{scalingformula}, 
consider, firstly, soliton field equations in 1+1 dimensions.

{\it Korteweg-de Vries equation.}
The KdV equation in physical form for scalar field $u(t,x)$ is given by
\EQ\label{kdveq}
\ueq{}{}(u,\Du{t},\Du{x},\Du{xxx}) 
= \Du{t} +u\Du{x} +\Du{xxx} =0 . 
\endEQ
This field equation is invariant under the scaling
$t\rightarrow \lambda^3 t$, $x\rightarrow \lambda x$, 
$u\rightarrow \lambda^{-2} u$. 
The corresponding scaling symmetry 
\EQ
\delta u=\scsymm{}{} = -2u-3t\Du{t}-x\Du{x}
\endEQ
is a solution of
the linearized field equation 
\EQ\label{kdvsymmeq}
\lin{\ueq{}{}}{\symm{}{}}{}{} 
= \D{t}\symm{}{} +\Du{x}\symm{}{} +u\D{x}\symm{}{} + \nD{x}{3}\symm{}{} =0
\endEQ
for all $u(t,x)$ satisfying the KdV equation. 
The adjoint of equation \eqref{kdvsymmeq} is given by 
\EQ\label{kdvadsymmeq}
\adlin{\ueq{}{}}{\adsymm{}}{}
= -\D{t}\adsymm{} -u\D{x}\adsymm{} - \nD{x}{3}\adsymm{} =0
\endEQ
whose solutions $\adsymm{}$ for all $u(t,x)$ satisfying \eqref{kdveq}
are the adjoint-symmetries of the KdV equation.
Here, we see $\adlinop{\ueq{}{}} \neq \linop{\ueq{}{}}$,
reflecting the fact that the KdV equation \eqref{kdveq} 
lacks a local Lagrangian 
formulation in terms of $u(t,x)$. 
Now, for any adjoint-symmetry $\adsymm{}$,
the conservation law formula \eqref{currformula} 
gives the conserved density
\EQ\label{kdvformula}
\P{t}{\adsymm{}} = -(3t\Du{t} +x\Du{x}+2u) \adsymm{} . 
\endEQ
From the obvious solution $w=u$ of equation \eqref{kdvadsymmeq},
we consider the infinite sequence of KdV adjoint-symmetries
$\adsymm{(k)} = (\adR)^k u$, $k=0,1,2,\ldots$, 
generated by the operator 
\EQ
\adR =\nD{x}{2} +\frac{1}{3}u +\frac{1}{3}\invDx(u\D{x})
\endEQ
which is the adjoint of the well-known KdV recursion operator \cite{Olver}
\EQ
\R=\nD{x}{2} +\frac{2}{3}u +\frac{1}{3}\Du{x}\invDx . 
\endEQ
Note that, under the KdV scaling symmetry, 
$\adsymm{(k)} \rightarrow \lambda^{-2(1+k)} \adsymm{(k)}$. 

Each adjoint-symmetry $\adsymm{(k)}$ is known to be a multiplier
for a local conservation law of the form 
\EQ\label{kdvconslaw}
\D{t} \curr{t}{(k)}(u,\Du{x},\Du{xx},\ldots)
+ \D{x} \curr{x}{(k)}(u,\Du{x},\Du{xx},\ldots)
=0
\endEQ
on KdV solutions $u(t,x)$,
through lengthy calculations. 
For instance, 
originally the KdV conservation laws were derived one by one via 
the Miura transformation \cite{KdV},
from which the multipliers can be calculated. 
An alternative approach has involved extracting 
the conserved densities one by one through a residue method 
using a formal symmetry (pseudo-differential operator) \cite{Olver-book} 
for the KdV equation. 
More recently, in \Ref{Anco-Bluman1}
the conserved densities were obtained one at a time 
from a homotopy integral formula
in terms of the adjoint-symmetries, 
after a verification of the adjoint invariance condition
on each one. 

Here, by-passing such cumbersome steps, 
formula \eqref{kdvformula} yields
the resulting conserved densities directly in terms of $\adsymm{(k)}$, 
\EQ\label{kdvcurrs}
\P{t}{\adsymm{}} 
\simeq 
\invDx(\Du{x}\adsymm{(k)}) -2u\adsymm{(k)} +3t(u\Du{x} +\Du{xxx})\adsymm{(k)}
\simeq 
-3\adR(\adsymm{(k)})
\endEQ
which follows by means of properties of $\adR$. 
This leads to a simple explicit recursion formula
for all of the KdV local conservation laws
\EQ\label{kdvscalingformula}
\curr{t}{(k)} = \frac{3}{3+2k} \adsymm{(k+1)}
= \frac{1+2k}{3+2k} \adR(\curr{t}{(k-1)}) , 
\endEQ
to within a trivial conserved density $\D{x}\triv{}$,
as a result of the scaling formula \eqref{scalingformula}.

{\it Sine-Gordon equation.}
The sine-Gordon equation is given by
\EQ\label{sgeq}
\ueq{}{}(u,\Du{tx}) 
= \Du{tx} -\sin u =0
\endEQ
for scalar field $u(t,x)$. 
This is a Lagrangian field equation
with the scaling invariance
$t\rightarrow \lambda^{-1} t$, $x\rightarrow \lambda x$, $u\rightarrow u$. 
The corresponding scaling symmetry is
$\delta u=\scsymm{}{} = t\Du{t}-x\Du{x}$ which is a solution of
the linearized field equation 
\EQ\label{sgsymmeq}
\lin{\ueq{}{}}{\symm{}{}}{}{} 
= \D{t}\D{x}\symm{}{} -(\cos u) \symm{}{} =0
\endEQ
for all $u(t,x)$ satisfying the sine-Gordon equation. 
Here, as we have $\adlinop{\ueq{}{}} = \linop{\ueq{}{}}$,
symmetries are the same as adjoint-symmetries, 
$\adsymm{}=\symm{}{}$. 
Hence, for any symmetry $\symm{}{}$,
the conservation law formula \eqref{currformula} 
gives the conserved density
\EQ\label{sgformula}
\P{t}{\symm{}{}} = (x\Du{x} -t\Du{t}) \D{x}\symm{}{} . 
\endEQ
We now consider the well-known infinite sequence of symmetries
$\symm{}{(k)} = (\R)^k \Du{x}$, $k=0,1,2,\ldots$, 
generated by the sine-Gordon recursion operator \cite{Olver}
\EQ
\R=\nD{x}{2} +\Du{x}\invDx(\Du{x}\D{x}) ,
\endEQ
starting from the translation symmetry $\symm{}{}= \Du{x}$.
Under the sine-Gordon scaling, note 
$\symm{}{(k)} \rightarrow \lambda^{-2k} \symm{}{(k)}$. 
Each symmetry $\symm{}{(k)}$ is a multiplier
for a local conservation law of the form \eqref{kdvconslaw}
on sine-Gordon solutions $u(t,x)$. 
These conservation laws were found originally 
by an application of Noether's theorem \cite{Dodd-Bullough}
and subsequently were derived by the same techniques 
used for the KdV equation \cite{Olver-book}, 
yielding the conserved densities one at a time
through lengthy calculations. 
Here, similarly to the KdV case,  
formula \eqref{sgformula} directly leads 
instead to a simple explicit expression
for all of the sine-Gordon local conservation laws, 
\EQ\label{sgconslaws}
\P{t}{\symm{}{}} \simeq 
-\invDx( \Du{x}\D{x}\symm{}{(k)} )
= - \nlR( \symm{}{(k)} )/\Du{x}
= \P{t}{(k)} 
\endEQ
and hence 
\EQ\label{sgscalingformula}
\curr{t}{(k)} \simeq \frac{-1}{1+2k} \P{t}{(k)}
= \frac{1}{1+2k} (\symm{}{(k+1)} -\nD{x}{2}\symm{}{(k)})/\Du{x}
\endEQ
from the scaling formula \eqref{scalingformula}.
Here $\nlR$ stands for the nonlocal part of $\R$. 
If the relation $-(\D{x}\P{t}{(k)})/\Du{x} = \D{x}\symm{}{(k)}$
is substituted into expression \eqref{sgscalingformula} to get 
\EQ
\symm{}{(k+1)} =
-\Du{x}\P{t}{(k)} -\D{x}( \frac{1}{\Du{x}}\D{x}\P{t}{(k)} )
\endEQ
then equation \eqref{sgconslaws}
yields an explicit conservation law recursion formula 
\EQ
\curr{t}{(k)} = \frac{2k-1}{2k+1} \PR( \curr{t}{(k-1)} )
\endEQ
with 
\EQ
\PR = 
\Du{x}{}^2\D{x}\frac{1}{\Du{x}{}^2}\D{x} 
+\frac{1}{2}\Du{x}{}^2 +\invDx( 
(\frac{\Du{xxx}}{\Du{x}} +\frac{1}{2}\Du{x}{}^2)\D{x} )
\endEQ
representing a recursion operator on conserved densities. 

{\it Modified Korteweg-de Vries vector equation.}
It is known that the recursion operator and symmetry hierarchy of
the sine-Gordon equation are closely related to that of 
the modified Korteweg-de Vries equation 
$\Du{t} +\frac{3}{2} u^2\Du{x} +\Du{xxx} =0$. 
This scalar field equation has an interesting generalization \cite{Wolf}
\EQ\label{mkdveq}
\vecueq{}(\vecu,\Dvecu{t},\Dvecu{x},\Dvecu{xxx}) 
= \Dvecu{t} +\frac{3}{2}\vecu{\cdot}\vecu\ \Dvecu{x} +\Dvecu{xxx} =0
\endEQ
for an $N$-dimensional vector field $\vecu(t,x)$, with any $N\geq 1$. 
The vector mKdV equation \eqref{mkdveq} is 
invariant under the scaling
$t\rightarrow \lambda^3 t$, $x\rightarrow \lambda x$, 
$\vecu\rightarrow \lambda^{-1} \vecu$. 
Its symmetries are the solutions of the linearized field equation
\EQ\label{mkdvsymmeq}
\lin{\vecueq{}}{\vecsymm{}}{}{} 
= \D{t}\vecsymm{} +3\vecu{\cdot}\vecsymm{}\ \Dvecu{x}
+\frac{3}{2}\vecu{\cdot}\vecu\  \D{x}\vecsymm{} 
+ \nD{x}{3}\vecsymm{} =0
\endEQ
for all mKdV solutions $\vecu(t,x)$, 
while its adjoint-symmetries 
are the solutions of the adjoint of equation \eqref{mkdvsymmeq} 
\EQ\label{mkdvadsymmeq}
\adlin{\vecueq{}}{\vecadsymm{}}{}
= -\D{t}\vecadsymm{} -\frac{3}{2}\vecu{\cdot}\vecu\ \D{x}\vecadsymm{} 
+3\Dvecu{x}{\cdot}\vecadsymm{}\ \vecu
-3\vecu{\cdot}\Dvecu{x}\ \vecadsymm{} 
- \nD{x}{3}\vecadsymm{} =0 . 
\endEQ
The vector mKdV equation \eqref{mkdveq}
admits the recursion operator 
\EQ\label{mkdvRop}
\R=
\nD{x}{2} +\vecu\cdot\vecu +\Dvecu{x}\invDx( \vecu\cdot\ )
-\vecu \intprod\invDx( \Dvecu{x}\wedge\ ) , 
\endEQ
where 
``$\wedge$'' denotes the antisymmetric outer product of two vectors
and ``$\intprod$'' denotes the interior product (\ie/ contraction) of
a vector against a tensor, 
namely ${\vec c}\intprod({\vec a}\wedge{\vec b}) = 
({\vec c}\cdot{\vec a}) {\vec b} - ({\vec c}\cdot{\vec b}) {\vec a}$. 
This expression \eqref{mkdvRop} is a manifestly $SO(N)$-invariant
version of the vector mKdV recursion operator first derived in \Ref{Wang}. 
The adjoint of $\R$ is given by the similar operator
\EQ
\adR=
\nD{x}{2} +\vecu \intprod(\vecu\wedge\ )
+ \Dvecu{x}\intprod \invDx( \vecu\wedge\ )
+ \vecu\invDx( \vecu\cdot\D{x}\ )
\endEQ
(closely resembling the form of the sine-Gordon recursion operator
in the scalar case $N=1$, when the ``$\wedge$'' terms vanish). 
We now consider the infinite sequence of mKdV adjoint-symmetries
generated by 
$\vecadsymm{(k)} = (\adR)^k \vecu$, $k=0,1,2,\ldots$, 
starting from the obvious solution of equation \eqref{mkdvadsymmeq}, 
$\vecadsymm{}=\vecu$. 
By means of the scaling symmetry 
\EQ
\delta \vecu=\vecscsymm{} = -\vecu-3t\Dvecu{t}-x\Dvecu{x} , 
\endEQ
the conservation law formula \eqref{currformula} 
yields the conserved density
\EQ\label{mkdvformula}
\P{t}{\adsymm{}} 
= -(3t\Dvecu{t} +x\Dvecu{x}+\vecu)\cdot \vecadsymm{}
\simeq
-\invDx(\vecu\cdot\D{x}\vecadsymm{(k)}) 
= \P{t}{(k)} . 
\endEQ
Note this can be expressed in terms of the recursions operator $\adR$ by
\EQ
\P{t}{(k)} = -\Dvecu{x}\cdot\nladR(\vecadsymm{(k)})/\vecu{\cdot}\Dvecu{x} . 
\endEQ
As $\vecadsymm{(k)} \rightarrow \lambda^{-(1+2k)} \vecadsymm{(k)}$
under the mKdV scaling symmetry, 
we see from the scaling formula \eqref{scalingformula} that 
each adjoint-symmetry $\vecadsymm{(k)}$ is a multiplier
for a local conservation law on mKdV solutions, given by 
\EQ\label{mkdvconslaw}
\D{t} \curr{t}{(k)}(\vecu,\Dvecu{x},\Dvecu{xx},\ldots)
+ \D{x} \curr{x}{(k)}(\vecu,\Dvecu{x},\Dvecu{xx},\ldots)
=0
\endEQ
with 
\EQ\label{mkdvscalingformula}
\curr{t}{(k)} \simeq \frac{-1}{2k+1} \P{t}{(k)} . 
\endEQ
However, in contrast to the scalar KdV and sine-Gordon cases, 
here expressions \eqrefs{mkdvformula}{mkdvscalingformula}
do not lead in any immediate way to a recursion operator
for the mKdV conservation laws
(since $\D{x}\vecadsymm{(k)}$ cannot be expressed directly in terms of 
$\P{t}{(k)}$, $\vecu$, and their derivatives). 
Nevertheless we have an explicit recursion formula
\EQ
\curr{t}{(k)} = \frac{1}{2k+1} \invDx(\vecu\cdot\D{x}( (\adR)^k \vecu ))
\endEQ
for generating all of the local conservation laws \eqref{mkdvconslaw}
(to within a trivial conserved density $\D{x}\triv{}$). 

Other soliton field equations, 
like the nonlinear Schrodinger equation, Tzetzeica equation, 
Harry-Dym equation, Boussinesq equation, 
and their variants \cite{solitoneqs}, 
as well as more general multi-component scalar/vector field equations 
\cite{Wolf},
can be treated in a similar way to the preceeding examples.

\subsection{ Fluid flow and wave propagation }

Secondly, field equations for fluid flow and nonlinear wave propagation
in 2+1 and 3+1 dimensions will be considered.

{\it Euler equations.}
The field equations for an incompressible inviscid fluid in two or three
spatial dimensions are given by 
\EQ\label{fluideq}
\fl\quad
\ueq{i}{}(\vecu,\der{t}\vecu,\vecder\vecu,P)
= \der{t}\u{i} +\u{j}\der{j} \u{i} +\invdens\coder{i}P =0 ,\quad
\ueq{}{}(\vecder\vecu)
= \der{i} \u{i} = 0
\endEQ
for fluid velocity $\u{i}(t,\vecx)$ and pressure $P(t,\vecx)$, 
with constant density $\dens$. 
This system is invariant under the family of scalings 
$t\rightarrow \lambda^p t$, $\x{i}{}\rightarrow \lambda \x{i}{}$, 
$\u{i}\rightarrow \lambda^{1-p} \u{i}$, $P\rightarrow \lambda^{2-2p} P$,
for arbitrary $p=\const$. 
The fluid symmetries are solutions $(\symm{}{i},\symm{}{})$ of
the linearized field equations
\EQ\label{fluidsymmeq}
\fl\quad
\lin{\vecueq}{\vecsymm{},\symm{}{}}{i}{} 
= \D{t}\symm{}{i} +\u{j}\D{j} \symm{}{i} +\der{j}\u{i}\symm{}{j} 
+\invdens\coD{i}\symm{}{} =0 ,\quad
\lin{\ueq{}{}}{\vecsymm{}}{i}{} =\D{i}\symm{}{i} =0
\endEQ
for all $\vecu(t,\vecx),P(t,\vecx)$ satisfying the Euler equations.
The adjoint of equations \eqref{fluidsymmeq} is given by 
\EQ\label{fluidadsymmeq}
\fl\quad
\adlin{\vecueq}{\vecadsymm{},\adsymm{}}{i}
= -\D{t}\adsymm{i} -\u{j}\D{j} \adsymm{i} +\der{i}\u{j}\adsymm{j} 
-\D{i} \adsymm{} =0 ,\quad
\adlin{\ueq{}{}}{\vecadsymm{}}{} =-\invdens\coD{i}\adsymm{i} =0
\endEQ
whose solutions $(\adsymm{i},\adsymm{})$ 
for all $\vecu(t,\vecx),P(t,\vecx)$ satisfying \eqref{fluideq}
are the adjoint-symmetries of the Euler equations.
In addition to scaling symmetries, 
\EQs
&&
\delta \u{i}=\scsymm{}{i} = (1-p)\u{i}-\x{j}{}\der{j}\u{i}-pt\der{t}\u{i} ,
\\&&
\delta P=\scsymm{}{} = (2-2p)P -\x{j}{}\der{j}P-pt\der{t}P ,
\endEQs
the Euler equations are well-known to possess 
the Galilean group \cite{Landau-Lifshitz,Olver-book} of symmetries, 
comprising time translations
\EQ\label{fluidtKVsymm}
\delta \u{i}=\symm{}{i} =\der{t}\u{i}
= -\u{j}\der{j} \u{i} -\invdens\coder{i}P ,\quad
\delta P=\symm{}{}= \der{t}P , 
\endEQ
Galilean boosts, with velocity $\v{i}=\const$, 
\EQ\label{fluidboostsymm}
\delta \u{i}=\symm{}{i} =t\v{j}\der{j}\u{i} -\v{i} ,\quad
\delta P=\symm{}{} = t\v{j}\der{j}P , 
\endEQ
and space translations and rotations
\EQ\label{fluidKVsymm}
\delta \u{i}=\symm{}{i} =\lieder{\kv{}{}}\u{i}
= \kv{j}{}\der{j}\u{i} -(\u{j}\der{j}\kv{i}{}) ,\quad
\delta P=\symm{}{} = \lieder{\kv{}{}}P 
= \kv{j}{}\der{j}P , 
\endEQ
where $\kv{i}{}(x)$ is a Killing vector of the Euclidean space 
in which the fluid flow takes place, \ie/ $\coder{(i}\kv{j)}{}=0$. 
(In particular, 
$\kv{i}{}=a^{i}=\const$ yields translations, 
and $\kv{i}{}=b^{ij}\x{}{j}$, $b^{ij}=-b^{ji}=\const$,
yields rotations.)
Corresponding adjoint-symmetries are given by the relations
\EQs
&&
\adsymm{i} =\parder{\symm{}{j}}{(\coder{i}\u{j})} ,
\\&&
\adsymm{} =\int \adsymm{i} d\u{i} -\der{t}\adsymm{i} \rmd\x{i}{}
+P \parder{\symm{}{i}}{(\coder{i}P)} ,
\endEQs
yielding 
\EQs
&&
\adsymm{i} =\id{ij}{}\u{j} ,\quad
\adsymm{} =\frac{1}{2}\id{ij}{}\u{i}\u{j} +\invdens P , 
\label{fluidtKVadsymm}\\
&&
\adsymm{i} =\id{ij}{}\kv{j}{} ,\quad
\adsymm{} =\id{ij}{}\kv{i}{}\u{j} , 
\label{fluidKVadsymm}\\
&&
\adsymm{i} =\id{ij}{} t\v{j} ,\quad
\adsymm{} =\id{ij}{} \v{i} (t\u{j} -\x{j}{}) .
\label{fluidboostadsymm}
\endEQs
Now, typically, 
local conservation laws 
\EQ
\D{t}\curr{t}{}(\vecx,\vecu,P) + \D{i}\curr{i}{}(\vecx,\vecu,P) =0
\endEQ
on solutions of the Euler equations
are derived through consideration of
Newton's laws applied to fluid elements \cite{Landau-Lifshitz}
or by Noether's theorem in a Hamiltonian formulation 
\cite{Olver-book,fluidnoether}.
In contrast, the conservation law formula \eqref{currformula} 
in terms of any adjoint-symmetry $(\adsymm{i},\adsymm{})$
directly yields a conserved density
\EQ\label{fluidformula}
\fl\quad
\P{t}{(\vecadsymm{},\adsymm{})}
=( (1-p)\u{i} -\x{j}{}\der{j}\u{i}-pt\der{t}\u{i} )\adsymm{i}
\simeq ( (pt\u{j} -\x{j}{})\der{j}\u{i} +(1-p)\u{i} )\adsymm{i}
\endEQ
to within a trivial conserved density $\D{i}\triv{i}$.
Here, this formula easily leads to 
momentum and angular momentum
\EQ\label{fluidmomentum}
\curr{t}{\veckv} = \id{ij}{} \kv{i}{} \u{j} 
\simeq \P{t}{\veckv}/\w{\veckv}
\endEQ
from the Killing vector adjoint-symmetries \eqref{fluidKVadsymm},
and Galilean momentum
\EQ\label{fluidgalileanmomentum}
\curr{t}{\vecv} = \id{ij}{} t\v{i} \u{j} 
\simeq \P{t}{\vecv}/\w{\vecv}
\endEQ
from the boost adjoint-symmetry \eqref{fluidboostadsymm},
as well as energy 
\EQ\label{fluidenergy}
\curr{t}{\vecu} = \frac{1}{2}\id{ij}{} \u{i} \u{j} 
\simeq \P{t}{\vecu}/\w{\vecu}
\endEQ
from the fluid velocity adjoint-symmetry \eqref{fluidtKVadsymm},
to within proportionality factors. 
(A useful identity in these calculations is 
$\u{j} = \D{i}( \x{j}{}\u{i} )$ on fluid solutions.)
From the scaling formula \eqref{scalingformula},
in three dimensions, it follows that 
$\w{\veckv}=4-p$ when $\veckv$ is a translation,
$\w{\veckv}=5-p$ when $\veckv$ is a rotation, 
$\w{\vecu}=5-2p$ and $\w{\vecv}=4$, 
representing the scaling weights of, respectively, 
the integrals for 
momentum and angular momentum 
$\int \veckv{\cdot}\vecu\ \rmd^3x$, 
energy $\int \frac{1}{2}|\vecu|^2 \rmd^3x$,
and Galilean momentum $\int t\vecv{\cdot}\vecu\ \rmd^3x$.
Note, for the dilation scaling $p=1$, all the scaling weights are positive
and hence the conservation laws 
\eqref{fluidmomentum}, \eqref{fluidgalileanmomentum}, \eqref{fluidenergy}
are noncritical. 
These weights decrease by $1$ in two dimensions,
leading to the same conclusions. 

The Euler equations also are known to possess 
a vorticity conservation law \cite{vorticity}, 
which is unrelated to symmetries in contrast with 
the energy and momentum conservation laws \cite{fluidnoether}. 
Here, a derivation will be given by formula \eqref{fluidformula}
directly in terms of fluid adjoint-symmetries. 

In three dimensions, 
vorticity is the curl of the fluid velocity
\EQ
\vort{i} = \cross{i}{jk} \coder{j}\u{k}
\endEQ
satisfying the vorticity equations \cite{Landau-Lifshitz}
\EQ
\der{t}\vort{i} +\u{j}\der{j}\vort{i} -\vort{j} \der{j}\u{i} =0 ,\quad
\der{i} \vort{i} =0 , 
\endEQ
where $\cross{i}{jk}$ is the cross-product operator
(\ie/ $\cross{}{ijk} = \id{li}{} \cross{l}{jk} = \cross{}{[ijk]}$
is the totally antisymmetric symbol).
We observe these equations have precisely the form of 
the adjoint-symmetry equations \eqref{fluidadsymmeq}
and hence
\EQ
\adsymm{i} = \id{ij}{} \vort{j} ,\quad 
\adsymm{} = 0
\endEQ
yields a corresponding fluid adjoint-symmetry. 
Then the conserved density formula \eqref{fluidformula} leads to
\EQ\label{fluidvorticity}
\curr{t}{\vecvort} = \frac{1}{2}\cross{}{ijk} \u{i} \coder{j} \u{k} 
\simeq \P{t}{\vecvort}/\w{\vecvort}
\endEQ
where, from scaling formula \eqref{scalingformula}, 
$\w{\vecvort}=4-2p$ is the scaling weight of the vorticity integral
$\int \vecu \cdot( \grad\times\vecu ) \rmd^3x$.
Physically, this conserved quantity describes the total helicity
(degree of knottedness) of vortex filaments. 
Note its scaling weight is noncritical provided $p\neq 2$. 

The situation in two dimensions is slightly different. 
The role of fluid vorticity is played by the scalar curl
\EQ
\vort{} = \cross{}{jk} \coder{j}\u{k}
\endEQ
where $\cross{}{jk} = \cross{}{[jk]}$ is the antisymmetric symbol. 
This scalar vorticity satisfies the conservation equation
\EQ\label{fluidvorteq}
\der{t}\vort{} +\u{j}\der{j}\vort{} =0 . 
\endEQ
By taking a curl, 
we immediately see that
\EQ
\adsymm{i} = \cross{}{ij}\coder{j} \vort{} ,\quad 
\adsymm{} = -\vort{2}/2
\endEQ
satisfy the fluid adjoint-symmetry equations \eqref{fluidadsymmeq}. 
The conserved density formula \eqref{fluidformula} now yields
\EQ\label{fluidvorticity'}
\curr{t}{\vort{}} =  \frac{1}{2} (\cross{}{jk} \coder{j} \u{k})^2 
\simeq \P{t}{\vort{}}/\w{\vort{}}
\endEQ
with the proportionality factor $\w{\vort{}}=2-2p$ 
given by the scaling weight of the conserved vorticity integral
$\int ( \grad\times\vecu )^2 \rmd^2x$
(where $\grad\times\vecu$ denotes the scalar curl 
$\cross{}{jk} \coder{j} \u{k}$). 
Note that this vorticity quantity is noncritical if $p\neq 1$,
\ie/ other than for a dilation scaling. 
More generally, 
any function $f(\vort{})$ in two dimensions is also a conserved density 
due to conservation \eqref{fluidvorteq} 
of the vorticity $\vort{}=\grad\times\vecu$, 
but the resulting  vorticity integral
$\int f(\grad\times\vecu) \rmd^2x$
will have a well-defined scaling weight only if $p=0$, 
namely for a spatial dilation scaling. 
In this case the conserved density
\EQ
\curr{t}{f} = f(\cross{}{jk} \coder{j} \u{k})
\simeq \P{t}{f}/\w{f}
\endEQ
again arises directly from formula \eqref{fluidformula},
through the adjoint-symmetry
\EQ
\adsymm{i} = \cross{}{ij} \coder{j}\vort{} f'' ,\quad
\adsymm{} = f-\vort{} f' . 
\endEQ
Here the proportionality factor is simply $\w{f}=2$,
due to the scaling invariance of $f(\vort{})$,
and consequently the vorticity integral is noncritical. 
The same conclusion holds for the vorticity integral 
in three dimensions if a spatial dilation scaling, $p=0$, 
is considered. 

The Navier-Stokes equations and polytropic gas dynamics equations
can be treated analogously.

{\it Nonlinear wave equation.}
The scalar field equation for $u(t,\vecx)$ 
\EQ\label{waveeq}
\ueq{}{}(u,\der{t}u,\der{\vecx}u)
= \nder{t}{2}u -\coder{i}\der{i}u \pm u^\upsilon 
= -\g{\alpha\beta}{} \der{\alpha}\der{\beta} u \pm u^\pow =0 
\endEQ
describes a nonlinear wave with interaction strength 
depending on a positive integer $\pow> 1$ 
(where $\g{}{\alpha\beta}$ is the Minkowski metric tensor
and $\x{\alpha}{}=(t,\vecx)$ are Minkowski spacetime coordinates). 
This is a Lagrangian wave equation 
invariant under the scaling
$\x{\alpha}{}\rightarrow \lambda \x{\alpha}{}$, 
$u\rightarrow \lambda^{q} u$, 
for $q=2/(1-\pow) \neq 0$. 
The corresponding scaling symmetry
\EQ
\delta u=\scsymm{}{} = q u -\x{\alpha}{}\der{\alpha}u
\endEQ
is a solution of the linearized field equation
\EQ\label{wavesymmeq}
\lin{\ueq{}{}}{\symm{}{}}{}{} 
= -\g{\alpha\beta}{} \D{\alpha}\D{\beta} u \pm \pow u^{\pow-1}\symm{}{} =0 
\endEQ
for all $u(t,\vecx)$ satisfying the wave equation \eqref{waveeq}.
Here, symmetries are the same as adjoint-symmetries, 
since $\adlinop{\ueq{}{}} = \linop{\ueq{}{}}$. 
The additional spacetime symmetries (Poincar\'e group) of 
the wave equation \eqref{waveeq}
for arbitrary $\pow>0$ are given by 
translations, rotations, and boosts,
\EQ\label{waveKVsymm}
\delta u=\symm{}{} =\lieder{\kv{}{}} u
= \kv{\alpha}{}\der{\alpha} u
\endEQ
where $\kv{\alpha}{}(x)$ is a Killing vector of Minkowski space, 
$\coder{(\alpha}\kv{\beta)}{}=0$. 
For certain interaction powers, 
$\pow=5$ in 2+1 dimensions and $\pow=3$ in 3+1 dimensions, 
the wave equation \eqref{waveeq} also admits \cite{Strauss}
inversion symmetries
\EQ\label{waveCKVsymm}
\delta u=\symm{}{} =\lieder{\kv{}{}} u +\frac{1}{6} \div\kv{}{}\ u
\endEQ
(where $\div\kv{}{}=\der{\alpha}\kv{\alpha}{}$)
associated with conformal Killing vectors
$\kv{\alpha}{} =
c^\beta \x{}{\beta} \x{\alpha}{} 
-\frac{1}{2} c^\alpha \x{}{\beta} \x{\beta}{}$,
$c^\beta=\const$, 
satisfying 
$\coder{(\alpha}\kv{\beta)}{} =\Omega\g{\alpha\beta}{}$
for a conformal factor $\Omega(x)$.
Through Noether's theorem, 
all spacetime symmetries \eqrefs{waveKVsymm}{waveCKVsymm}
are known to be multipliers \cite{Strauss}
for local conservation laws 
on solutions of the wave equation \eqref{waveeq},
\EQ
\D{\alpha}\curr{\alpha}{\kv{}{}}(t,\vecx,u,\der{t}u,\der{\vecx}u)=0
\endEQ
with 
\EQ
\curr{\alpha}{\kv{}{}} =
- \T{\alpha}{\beta}(u,\ujet{})\kv{\beta}{}
-\frac{1}{\pow+1}( \div\kv{}{}\, u\coder{\alpha}u 
-\frac{1}{2} (\coder{\alpha}\div\kv{}{}) u^2 )
\endEQ
given in terms of the conserved stress-energy tensor 
\EQ\label{waveT}
\T{\alpha}{\beta}(u,\ujet{}) =
\coder{\alpha}u\der{\beta}u 
-\frac{1}{2}\id{\beta}{\alpha}( \coder{\gamma}u\der{\gamma}u 
\pm \frac{2}{\pow+1} u^{\pow+1} ) . 
\endEQ
The conservation $\der{\alpha}\T{\alpha}{\beta}(u,\ujet{})=0$ of 
this tensor \eqref{waveT}
on solutions of the wave equation
provides a well-known alternative derivation of the conservation laws
associated with spacetime Killing vectors, 
\EQ
\curr{\alpha}{\kv{}{}} =
- \T{\alpha}{\beta}(u,\ujet{})\kv{\beta}{}
\quad\eqtext{for $\coder{(\alpha}\kv{\beta)}{}=0$.}
\endEQ
Here, the resulting conserved densities instead will be obtained 
from the conservation law formula \eqref{currformula} 
directly in terms of the corresponding symmetries 
\eqrefs{waveKVsymm}{waveCKVsymm}. 
This yields
\EQs
\fl\quad
\P{\alpha}{\symm{}{}} &&
= ( (q-1)\coder{\alpha}u -\x{\beta}{}\coder{\alpha}\der{\beta}u )
( \kv{\gamma}{}\der{\gamma}u +\frac{1}{\pow+1}\div\kv{}{}\ u)
\nonumber\\\fl\quad&&\qquad
- ( q u - \x{\beta}{}\der{\beta}u )
( \coder{[\alpha}\kv{\gamma]}{}\der{\gamma}u 
+ \kv{\gamma}{}\coder{\alpha}\der{\gamma}u 
+ \frac{1}{2} \div\kv{}{} \coder{\alpha}u 
+\frac{1}{\pow+1} (\coder{\alpha}\div\kv{}{}) u )
\nonumber\\\fl\quad&&
\simeq -\w{\bdkv} \curr{\alpha}{\kv{}{}}
\label{waveformula}
\endEQs
to within a trivial conserved density
($\D{\beta}\triv{\alpha\beta}$, 
with $\triv{\alpha\beta}=-\triv{\beta\alpha}$). 
The proportionality factor $\w{\bdkv}$ is, 
by the scaling formula \eqref{scalingformula},
the scaling weight of the flux integrals
$\int \curr{\alpha}{\kv{}{}} \n{}{\alpha} \rmd\Sigma$, 
on a $t=\const$ spatial hypersurface $\Sigma$ 
with normal vector $\n{}{\alpha}=\der{\alpha}t$. 
For translations 
$\kv{\alpha}{}=a^{\alpha}=\const$, 
we have 
\EQ
\w{\bdkv}=4/(1-\pow)
\endEQ
in 2+1 dimensions, 
while in 3+1 dimensions, 
\EQ
\w{\bdkv}=(5-\pow)/(1-\pow) ; 
\endEQ
the weight $\w{\bdkv}$ increases by $1$ for rotations and boosts 
$\kv{\alpha}{}=b^{\alpha\beta}\x{}{\beta}$, 
$b^{\alpha\beta}=-b^{\beta\alpha}=\const$,
and increases by $1$ again for inversions \eqref{waveCKVsymm}
so thus $\w{\bdkv}=1$ in the case of proper conformal Killing vectors. 
Hence, a critical case $\w{\bdkv}=0$ only occurs 
for translations in 3+1 dimensions when $\pow=5$, 
and for rotations and boosts in 3+1 dimensions when $\pow=3$
as well as in 2+1 dimensions when $\pow=5$, 
corresponding to scaling invariance of the energy-momentum integral
$\int -\T{\alpha}{\beta}(u,\ujet{})a^\beta \n{}{\alpha} \rmd\Sigma$
and of the angular-boost momentum integral 
$\int -\T{\alpha}{\beta}(u,\ujet{})b^{\beta\gamma}\x{}{\gamma} 
\n{}{\alpha} \rmd\Sigma$
in these cases. 
Consequently, $\P{\alpha}{\symm{}{}} \simeq 0$ is trivial
only for these critical interaction powers \cite{critical} 
and Killing vectors. 
In this situation, 
all local conservation laws are produced nevertheless
from the more general formula \eqref{curr}
directly in terms of pairs of 
Killing vector symmetries \eqref{waveKVsymm},
\EQ
\curr{\alpha}{}(\eta_1,\eta_2) 
\simeq 
-\T{\alpha}{\beta}(u,\ujet{})\kv{\beta}{}
\endEQ
where $\kv{\alpha}{} =[\kv{}{1},\kv{}{2}]^\alpha$
is the commutator of the Killing vectors.
The same result holds even for 
the non-critical cases where $\w{\bdkv}\neq 0$. 

Other nonlinear wave equations,
such as sigma models and wavemap equations \cite{Shatah}, 
can be treated in the same way. 

\subsection{ Gauge theories }

Finally, Yang-Mills fields and gravitational fields 
in 3+1 dimensions will be considered. 

{\it Yang-Mills theory.}
The Yang-Mills field on Minkowski space $(\Rnum^4,\g{}{\alpha\beta})$
is a vector potential $\A{a}{\alpha}(x)$ that takes values in 
an internal Lie algebra $\liealg=(\Rnum^N,\c{a}{bc})$. 
Associated with $\A{a}{\alpha}$ is the Yang-Mills covariant derivative
\EQ
\ymder{\alpha} = \der{\alpha} +\c{a}{bc} \A{b}{\alpha}
\endEQ
and the Yang-Mills field strength tensor
\EQ
\F{a}{\alpha\beta} = 
\der{[\alpha}\A{a}{\beta]} +\frac{1}{2}\c{a}{bc} \A{b}{\alpha}\A{c}{\beta}
\endEQ
where $\c{a}{bc}$ denotes 
the structure constants of the Lie algebra $\liealg$. 
The Yang-Mills equation with gauge group based on $\liealg$ 
is then given by 
\EQ\label{ymeq}
\ueq{a}{\mu}(A,\Ajet{},\Ajet{2}) 
= \g{\alpha\beta}{}\ymder{\alpha} \F{a}{\beta\mu} =0
\endEQ
which is invariant under the scaling 
$\x{\alpha}{} \rightarrow \lambda\x{\alpha}{}$, 
$\A{a}{\alpha} \rightarrow \lambda^{-1}\A{a}{\alpha}$. 
Whenever the gauge group is semisimple, 
the Yang-Mills equation \eqref{ymeq} arises from a Lagrangian 
(see \eg/ \Ref{Anco1}),
and in this situation 
both the symmetries and adjoint-symmetries of this field equation \eqref{ymeq}
are given by solutions $\delta\A{a}{\alpha} = \symm{\alpha}{a}$ of 
the linearized Yang-Mills equation
\EQ
\lin{\ueq{}{}}{\symm{}{}}{a}{\mu} = 
\g{\alpha\beta}{}( \ymder{\alpha}\ymder{[\beta}\symm{\mu]}{a}
+ \c{a}{bc} \symm{\alpha}{b} \F{c}{\beta\mu} ) =0
\endEQ
for all Yang-Mills solutions $\A{a}{\alpha}(x)$. 
Note, here, 
$\ymder{\alpha} = \D{\alpha} +\c{a}{bc} \A{b}{\alpha}$
acts as a total derivative operator. 
The well-known local symmetries of the Yang-Mills equation \eqref{ymeq} 
are comprised by 
gauge symmetries
\EQ\label{ymgaugesymm}
\delta\A{a}{\alpha} = \ymder{\alpha} \sf{a}
\endEQ
involving any Lie-algebra valued scalar function 
$\sf{a}(x,A,\Ajet{},\ldots)$,
and spacetime symmetries
\EQ\label{ymkvsymm}
\delta\A{a}{\alpha} = 2\kv{\beta}{} \F{a}{\beta\alpha}
= \lieder{\kv{}{}} \A{a}{\alpha} - \ymder{\alpha}( \kv{\beta}{}\A{a}{\beta} )
\endEQ
where $\kv{\beta}{}(x)$ is any conformal Killing vector on Minkowski space,
$\coder{(\alpha}\kv{\beta)}{} =\frac{1}{4}\g{\alpha\beta}{} \div\kv{}{}$. 
In the case of a dilation Killing vector, $\kv{\beta}{}=\x{\beta}{}$,
the spacetime symmetry \eqref{ymkvsymm} reduces to a sum of 
the Yang-Mills scaling symmetry and a gauge symmetry,
\EQ\label{ymscsymm}
\scvar\A{a}{\alpha} = \scsymm{\alpha}{a} 
= -\x{\beta}{}\der{\beta} \A{a}{\alpha} -  \A{a}{\alpha}
= -2\x{\beta}{} \F{a}{\beta\alpha} 
- \ymder{\alpha}( \x{\beta}{}\A{a}{\beta} ) . 
\endEQ
A recent classification analysis \cite{Pohjanpelto}
has proved that these are in fact the only nontrivial local symmetries
admitted by the Yang-Mills equation 
if the Lie algebra $\liealg$ is real and simple. 
However, for a simple Lie algebra $\liealg$ with a complex structure,
the same analysis found that the Yang-Mills equation also admits
complexified spacetime symmetries
\EQ\label{ymcomplexkvsymm}
\delta\A{a}{\alpha} = 2\j{a}{b}\kv{\beta}{} \F{b}{\beta\alpha}
\endEQ
where $\j{a}{b}$ is the complex structure map on $\liealg$,
satisfying the properties 
\EQ
\j{a}{b} \j{b}{c} = -\id{a}{c} ,\quad
\j{a}{b} \c{b}{cd} = \c{a}{ed} \j{e}{c} ,\quad
\ck{}{a[b} \j{a}{c]} =0
\endEQ
with  $\ck{}{ab} = \c{c}{ad} \c{d}{bc}$ 
being the Cartan-Killing metric on $\liealg$. 
Through Noether's theorem, 
these symmetries \eqsref{ymgaugesymm}{ymcomplexkvsymm} 
are multipliers for local conservation laws 
$\D{\alpha} \curr{\alpha}{}(x,A,\Ajet{}) =0$
consisting of, respectively, 
\EQ
\curr{\alpha}{} = 
\g{\alpha\nu}{}\g{\beta\gamma}{} \ck{}{ab} 
\F{a}{\nu\beta} \ymder{\gamma} \sf{b} 
\simeq 0
\endEQ
related to the Bianchi identity on $\F{a}{\alpha\beta}$, 
and 
\EQ\label{ymkvconslaw}
\curr{\alpha}{} = \T{\alpha}{\beta}(F) \kv{\beta}{} ,\quad
\curr{\alpha}{} = \jT{\alpha}{\beta}(F) \kv{\beta}{} ,
\endEQ
given by the conserved Yang-Mills stress-energy tensor
\EQ
\T{}{\alpha\beta}(F) = 
\g{\nu\sigma}{}\ck{}{ab}( \F{a}{\alpha\nu} \F{b}{\beta\sigma}
-\frac{1}{4}\g{}{\alpha\beta}\g{\mu\gamma}{} 
\F{a}{\mu\nu} \F{b}{\gamma\sigma} ) 
\endEQ
and its complexification
\EQ
\jT{}{\alpha\beta}(F) = 
\g{\nu\sigma}{}\j{}{ab}( \F{a}{\alpha\nu} \F{b}{\beta\sigma}
-\frac{1}{4}\g{}{\alpha\beta}\g{\mu\gamma}{} 
\F{a}{\mu\nu} \F{b}{\gamma\sigma} ) . 
\endEQ
These conservation laws \eqref{ymkvconslaw} yield (complexified)
energy-momentum, angular and boost momentum 
for translation, rotation and boost Killing vectors 
$\coder{(\alpha}\kv{\beta)}{} =0$,
and additional quantities for dilation and inversion Killing vectors
$\coder{(\alpha}\kv{\beta)}{} 
=\frac{1}{4}\div\kv{}{}\g{\alpha\beta}{} \neq 0$. 

However, the previous results do not fully settle the classification of
local conservation laws of the Yang-Mills equation \eqref{ymeq}, 
since it leaves open the question of whether any trivial symmetries
could yield nontrivial conservation laws by Noether's theorem. 
For Lagrangian field equations whose principal part 
(\ie/ highest derivative terms) is nondegenerate,
it is known that there is a one-to-one correspondence between 
nontrivial variational symmetries and nontrivial conservation laws
\cite{Olver-book,Anco-Bluman2}. 
But this correspondence is not automatic 
for a field equation with gauge symmetries, 
due to the resulting degeneracy of the field equation's principal part,
in contrast to the previous examples in this section. 
Here, through an application of formula \eqref{currformula}, 
the gap in the classification of Yang-Mills conservation laws
will be addressed. 

We consider local symmetries
\EQ\label{ymsymm}
\delta\A{a}{\alpha} = \symm{\alpha}{a}(x,A,\Ajet{},\ldots)
\endEQ
assumed to be homogeneous with respect to 
the Yang-Mills scaling \eqref{ymscsymm},
\EQ\label{ymsymmweight}
\scvar\symm{\alpha}{a} = 
r\symm{\alpha}{a} -\x{\beta}{}\der{\beta}\symm{\alpha}{a}
\endEQ
with scaling weight $r=\const$. 
As noted in \secref{formula},
there is no loss of generality in such a homogeneity restriction. 
Now, formula \eqref{currformula} yields a conserved current 
generated from any such symmetry, 
\EQ
\P{\alpha}{}(\symm{}{},\scvar A) \simeq
2\ck{ab}{}( 
\x{\mu}{}\F{a}{\mu\beta} \g{\beta[\nu}{}\ymcoder{\alpha]} \symm{\nu}{b}
- \symm{\nu}{a}( \x{\mu}{} \ymder{\mu} \F{b\alpha\nu}{} + \F{b\alpha\nu}{} ) 
) ,
\endEQ
with the current being linear and homogeneous 
in $\symm{\alpha}{a}$ and $\ymder{\mu}\symm{\alpha}{a}$. 
If $\symm{\alpha}{a}=\Q{a}{\alpha}$ is a multiplier for 
a local conservation law of the Yang-Mills equation \eqref{ymeq},
\EQ\label{ymconslaw}
\ck{}{ab} \g{\alpha\beta}{} \Q{a}{\alpha} 
\ymcoder{\nu} \F{b}{\nu\beta} 
= \D{\alpha} \curr{\alpha}{\Q{}{}} 
\endEQ
where the current 
\EQ
\curr{\alpha}{\Q{}{}}(x,A,\Ajet{},\ldots) 
\endEQ
can be assumed homogeneous under the Yang-Mills scaling \eqref{ymscsymm}, 
then the scaling formula \eqref{scalingformula} gives the relation
\EQ
\P{\alpha}{}(\symm{}{},\scvar A) \simeq
(r+1) \curr{\alpha}{\Q{}{}}
\endEQ
to within a trivial conserved current. 
Hence, for a variational symmetry that is trivial, 
so $\symm{\alpha}{a} =0$ for solutions of \eqref{ymeq}, 
we see that, if $r\neq -1$, 
\EQ
\curr{\alpha}{\Q{}{}} \simeq \frac{1}{r+1} \P{\alpha}{}(0,\scvar A) =0
\endEQ
is a trivial noncritical current on Yang-Mills solutions. 
Note the factor $r+1$ here is precisely the scaling weight of
the flux integral of the current $\curr{\alpha}{\Q{}{}}$. 
This establishes the following classification result. 

{\bf Proposition 3.1}: 
No nontrivial conservation laws \eqref{ymconslaw}
that are noncritical ---
\ie/ whose associated conserved quantity 
$\int \curr{\alpha}{} \der{\alpha}t \rmd\Sigma$, 
on a spatial hypersurface $\Sigma$ given by $t=\const$, 
has non-zero scaling weight ---
may arise from variational trivial symmetries 
\eqrefs{ymsymm}{ymsymmweight}
of the Yang-Mills equation \eqref{ymeq}. 

{\it General Relativity.}
The vacuum gravitational field equation \cite{Wald} 
on a 4-dimensional spacetime manifold is given by 
\EQ\label{greq}
\ueq{\alpha\beta}{}(g,\gjet{},\gjet{2}) =\G{}{\alpha\beta}=0
\endEQ
where $\G{\alpha\beta}{}$ is the Einstein tensor
(\ie/ trace-reversed Ricci tensor $\curv{\alpha\beta}{}$)
for the spacetime metric $\g{}{\alpha\beta}(x)$. 
Its linearized field equation is given by
\EQ\label{grsymmeq}
\lin{\ueq{}{}}{\symm{}{}}{\alpha\beta}{}
= -\frac{1}{2}( \covcoder{\mu}\covder{\mu} \trrvsymm{\alpha\beta}
-2\covder{\mu}\covcoder{(\alpha} \trrvsymm{\beta)\mu}
+ \g{\alpha\beta}{} \covder{\mu}\covder{\nu} \trrvsymm{\mu\nu} ) =0
\endEQ
whose solutions for all $\g{}{\alpha\beta}(x)$ 
satisfying the Einstein equation \eqref{greq}
are the symmetries 
$\delta\g{\alpha\beta}{}=\symm{}{\alpha\beta}(x,g,\gjet{},\ldots)$
of the gravitational field, 
where $\covder{\mu}$ is the covariant total derivative operator
associated with the metric, 
$\covder{\mu}\g{}{\alpha\beta}=0$, 
and ``bar'' denotes trace-reversal on a given tensor. 
Here, since the gravitational field equation \eqref{greq} 
comes from a Lagrangian \cite{Wald}, 
adjoint-symmetries are the same as symmetries. 
It is known that the only admitted nontrivial local symmetries 
\cite{Anderson-Torre} consist of 
a constant conformal scaling
\EQ\label{grscaling}
\delta \g{}{\alpha\beta}=\g{}{\alpha\beta}
\endEQ
and diffeomorphism gauge symmetries
\EQ\label{grdiffeo}
\delta \g{}{\alpha\beta}=\lieder{\vf{}{}}\g{}{\alpha\beta}
= 2\covder{(\alpha} \vf{}{\beta)}
\endEQ
for any local vector field $\vf{\beta}{}(x,g,\gjet{},\ldots)$. 
If we consider local conservation laws of 
the gravitational field equation \eqref{greq}, 
\EQ\label{grconslaw}
\covder{\alpha}\curr{\alpha}{}(x,g,\gjet{},\ldots) =0 , 
\endEQ
then through Noether's theorem 
the constant conformal scaling \eqref{grscaling} is not a multiplier,
while the diffeomorphisms \eqref{grdiffeo} yield 
a trivial conservation law 
$\curr{\alpha}{}=\vf{\beta}{} \G{\beta}{\alpha} =0$
on solutions of \eqref{greq}.

\Ref{Anderson-Torre} asserts that in fact 
there exist no nontrivial local conservation laws \eqref{grconslaw}, 
but does not provide a full statement of the proof. 
Here a complete classification proof will be given
for diffeomorphism-covariant conservation laws,
by use of a covariant version of 
the conservation law formula \eqref{currformula}.
A suitable dilation scaling symmetry is provided by 
the diffeomorphism symmetry \eqref{grdiffeo} 
specialized to a homothetic vector field $\vf{\alpha}{}=\scvf{\alpha}{}$,
namely 
\EQ\label{grhomothetic}
\scvar\g{\alpha\beta}{}=\lieder{\scvf{}{}}\g{}{} ,\quad
\covder{\alpha}\scvf{\alpha}{}=\const \neq 0 . 
\endEQ
Now, a local conservation law \eqref{grconslaw} 
is diffeomorphism-covariant whenever it is of the form 
\EQ\label{grnaturalconslaw}
\covder{\alpha}\curr{\alpha}{}(g,R,\covder{}R,\ldots)=0
\endEQ
on solutions of the Einstein equation \eqref{greq},
with $\curr{\alpha}{}$ satisfying the natural transformation property
$\delta\curr{\alpha}{}= \lieder{\vf{}{}}\curr{\alpha}{}$
under all diffeomorphism symmetries \eqref{grdiffeo}.
Correspondingly, without loss of generality
we consider diffeomorphism-covariant local symmetries 
\EQ\label{grnaturalsymm}
\delta\g{\alpha\beta}{}=\symm{}{\alpha\beta}(g,R,\covder{}R,\ldots) , 
\endEQ
which are necessarily homogeneous with respect to the dilation scaling
\EQ
\scvar\symm{}{\alpha\beta} = \lieder{\scvf{}{}}\symm{}{\alpha\beta} . 
\endEQ
Then for any such symmetry \eqref{grnaturalsymm}, 
formula \eqref{currformula} yields a conserved current
\EQ
\fl\quad
\P{\alpha}{}(\symm{}{},\lieder{\scvf{}{}}g) =
- ( \scvf{\nu}{} \curv{\nu\mu\beta}{\alpha}
- \covder{\mu} \covder{\beta} \scvf{\alpha}{} ) \trrvsymm{\mu\beta}
+ \covder{\mu}\scvf{}{\beta} \covcoder{\alpha} \trrvsymm{\mu\beta}
+ \covder{\mu}\scvf{\mu}{} \covder{\nu} \trrvsymm{\alpha\nu}
\endEQ
whose dependence on 
$\symm{}{\alpha\beta}$ and $\covder{\mu}\symm{}{\alpha\beta}$
is linear, homogeneous. 
It now follows from the scaling formula \eqref{scalingformula}
in covariant form that 
if $\symm{}{\alpha\beta}=\coQ{\alpha\beta}$ is a multiplier
for a local conserved current, 
$\Q{}{\alpha\beta}\G{}{\alpha\beta} = \covder{\alpha}\curr{\alpha}{\Q{}{}}$,
then to within a trivial conserved current,
\EQ\label{grformula}
\P{\alpha}{}(\symm{}{},\lieder{\scvf{}{}}g) \simeq 
\w{} \curr{\alpha}{\Q{}{}} ,\quad 
w=\covder{\beta}\scvf{\beta}{}\neq 0 . 
\endEQ
Hence, for a trivial multiplier
that is diffeomorphism-covariant \eqref{grnaturalsymm}, 
we have 
\EQ
\curr{\alpha}{\Q{}{}} \simeq 
\w{}\inv \P{\alpha}{}(0,\lieder{\scvf{}{}}g) =0
\endEQ
on solutions of the Einstein equation \eqref{greq}, 
since $\symm{}{\alpha\beta}=0$. 
Therefore, the following result holds. 

{\bf Proposition 3.2}: 
No nontrivial diffeomorphism-covariant conservation laws 
\eqref{grnaturalconslaw}
may arise from variational trivial symmetries 
\eqref{grnaturalsymm}
of the gravitational field equation \eqref{greq}. 

Solutions of the Einstein equation \eqref{greq}
with a Killing vector 
$\lieder{\kv{}{}}\g{}{\alpha\beta} = 2\covder{(\alpha} \kv{}{\beta)}=0$
are known to possess an unexpected extra symmetry \cite{Geroch1}
\EQ\label{grkvsymm}
\delta\g{}{\alpha\beta} = 
\twist\g{}{\alpha\beta} -2\kv{}{(\alpha} \dualkv{}{\beta)} ,\quad
\delta\kv{}{\alpha} = - \dualkv{}{\alpha} \kv{\beta}{}\kv{}{\beta} , 
\endEQ
where $\twist$ is the scalar twist 
and $\dualkv{}{\beta}$ is the dual of $\kv{\beta}{}$, 
defined by the equations
\EQ
\covder{\alpha}\twist 
=\vol{\alpha\beta\mu\nu}{} \kv{\beta}{}\covcoder{\mu}\kv{\nu}{} ,\quad
2\covder{[\alpha}\dualkv{}{\beta]} 
= \vol{\alpha\beta\mu\nu}{}\covcoder{\mu}\kv{\nu}{} .
\endEQ
Note this symmetry \eqref{grkvsymm} is nonlocal 
due to its dependence on $\twist,\dualkv{}{\beta}$
(in terms of $\g{}{\alpha\beta},\kv{}{\alpha}$).
Here, a corresponding nonlocal conservation law will be derived
through the conservation law formula \eqref{currformula} 
in covariant form
using the constant conformal scaling \eqref{grscaling},
\EQ\label{grcurr}
\P{\alpha}{}(\symm{}{},g) 
= \g{}{\nu\mu} \covcoder{\alpha} \symm{}{\nu\mu}
- \covder{\mu} \symm{}{\alpha\mu} . 
\endEQ
This yields, from \eqref{grkvsymm},
\EQ
\P{\alpha}{}(\symm{}{},g)
= \covcoder{\alpha}\twist +2\covder{\beta}( \kv{(\beta}{}\dualkv{\alpha)}{} )
\simeq 2 \kv{}{\beta} \covcoder{(\beta}\dualkv{\alpha)}{} . 
\endEQ
Moreover, 
on solutions of the Einstein equation \eqref{greq}
with two commuting Killing vectors 
$\kv{\alpha}{1}$ and $\kv{\alpha}{2}$, 
the symmetries \eqref{grkvsymm} are known to generate
an infinite-dimensional algebra of nonlocal symmetries 
(the Geroch group \cite{Geroch2}). 
Then the covariant formula \eqref{grcurr} leads to a corresponding 
infinite sequence of conservation laws,
related to the complete integrability of the system
$\G{\alpha\beta}{} =0$, 
$\lieder{\kv{}{1}}\g{}{\alpha\beta} = \lieder{\kv{}{2}}\g{}{\alpha\beta} =0$. 

A similar classification treatment of 
local and nonlocal conservation laws of 
the self-dual Einstein equation and self-dual Yang-Mills equation
will be given elsewhere.

\section{ Conclusion }
\label{conclude}

The conservation law expression derived in Proposition~2.1 
is a generalization of
a formula mentioned in \Ref{Olver-book} in the case of linear PDEs
and extends a similar formula considered for self-adjoint PDEs 
in \Ref{Anco-Bluman4}. 
A variant of this expression has been central to a recently obtained
classification of local conserved currents for 
linear massless spinorial field equations of spin $s>0$ 
\cite{maxwell,spins}. 
Numerous applications to nonlinear ODEs are presented in \Ref{ourbook}.

Apart from its main use in generating all noncritical local conservation laws
in an algebraic manner ---
by-passing the standard homotopy integral formula
and (adjoint-) invariance conditions ---
in terms of local (adjoint-) symmetries
admitted by any given PDEs, 
the conservation law expression is able to produce 
nonlocal conservation laws from any admitted nonlocal (adjoint-) symmetries. 
Such examples and applications will be considered in a forthcoming paper
\cite{nonlocal}.

\end{document}